\documentclass[11pt]{article}
\begin{document}

\title{\textbf{\textsf{Charged Black Holes in Phantom Cosmology}}}
\author{ Mubasher Jamil\footnote{mjamil@camp.edu.pk}, Muneer Ahmad
Rashid\footnote{muneerrshd@yahoo.com}\ \ and Asghar
Qadir\footnote{aqadirmath@yahoo.com}
\\ \\
\textit{\small Center for Advanced Mathematics and Physics}\\
\textit{\small National University of Sciences and Technology}\\
\textit{\small Peshawar Road, Rawalpindi, 46000, Pakistan} \\
} \maketitle
\begin{abstract}
In the classical relativistic regime, the accretion of phantom-like
dark energy onto a stationary black hole reduces the mass of black
hole. Here we have investigated the accretion of phantom energy onto
a stationary charged black hole and have determined the condition
under which this accretion is possible. This condition restricts the
mass to charge ratio in a narrow limit. This condition also
challenges the validity of the cosmic censorship conjecture since a
naked singularity is eventually produced as magnitude of charge
increases compared to mass of black hole.
\end{abstract}


\section{Introduction}

Accelerated expansion of the universe has been observed and
confirmed by myriad of sources including analysis of cosmic
microwave background radiation \cite{sper1}, large scale structure
\cite{eisen} and supernovae SNe Ia data \cite{perl,riess}. This
expansion is supposedly driven by exotic vacuum energy having $\rho
>0$ and $p<0$ or equivalently $p=-\rho $ (or $\omega =-1$),
dominating the observable universe. Observations of WMAP data
suggest that its magnitude is more than 70\% of the total\ energy
density of the universe \cite{sper}. Among other forms of\ exotic\
energies (e.g quintessence, cosmological constant, k-essence,
hessence etc), the phantom energy with $\omega <-1$ exhibits similar
behavior on large cosmic scale. The genesis of phantom energy is not
clear but it violates the null and weak energy conditions. As these
conditions are the weaker one, they ensure that the stronger
conditions (i.e. strong and dominant) will be violated automatically
\cite{johri,lobo,lobo1}. These energy conditions guarantee the
positive definiteness of the energy densities and pressure densities
of all the matter content in the universe. Recent observational data
constrain the range of dark energy by $-1.38<\omega<-0.82$ at 95\%
confidence level \cite{melch}. Rather the supernovae data favor an
evolving $\omega(z)$ varying from quintessence ($\omega>-1$) to
phantom regime($\omega<-1$) \cite{alam}. Further, the extrapolation
of WMAP data is best fitted with the notion of phantom energy
\cite{cald}

The energy density and the pressure of the phantom energy can be
represented by the minimally coupled spatially homogeneous and time
dependent scalar field $\phi$ having negative kinetic energy term
given by
\begin{equation}
\rho=-\frac{\dot{\phi}^2}{2}+V(\phi),\ \
p=-\frac{\dot{\phi}^2}{2}-V(\phi).
\end{equation}
Here $V(\phi)$ is the scalar potential and dot over $\phi$
represents the derivative with respect to time parameter $t$. Note
that if the kinetic term in Eq. (1) is positive then it gives usual
dark energy with satisfies all the energy conditions. The above
parameters $\rho$ and $p$ are related to the Hubble parameter $H$ as
\begin{equation}
H^2=\frac{4\pi}{3}(-\dot{\phi}^2+2V),
\end{equation}
where $H(t)=\dot{a}/a$ and $a(t)$ is the scale factor which arises
in the Friedmann-Robertson-Walker spacetime. From Eq. (2), we
require the potential $V(\phi)$ to be positive. It is argued by
using scalar field models of phantom energy, that it can behave as a
long range repulsive force \cite{Luca}. The phantom energy possesses
some peculiar properties unlike normal matter e.g. (1) its energy
density $\rho(t)$ increases with the expansion of the universe, (2)
it ensures the existence and stability of traversable worm holes in
the universe \cite{bronn3,ellis,picon,rahm,kuhf}, (3) also
self-gravitating, static and spherically symmetric phantom scalar
fields with arbitrary potentials can generate a stable configuration
of a regular black hole or apparently non-singular black hole which
inherently possesses exactly Schwarzschild-like causal structure but
the singularity is replaced by a de Sitter infinity, thereby
generating an asymptotically de Sitter expansion beyond the black
hole horizon \cite{bronn1,bronn2}, (4) due to strong negative
pressure the phantom energy can disrupt all gravitationally bound
structures i.e from galactic clusters to the gravitationally
collapsed objects including black holes
\cite{babi,Ness,mota1,mota2,babi2,babi3}, (5) it can produce
infinite expansion of the universe in a finite time thus causing the
`big rip' (i.e. a state when $a(t), \rho(t)\rightarrow\infty$ for
$t<\infty$) \cite{johri,cald}.

The big rip is characterized by a future singularity implying a
finite age of the universe. It has been proposed that this future
singularity can be avoided if the phantom energy is interacting with
the dark matter \cite{curbelo,nojiri}. The interaction of phantom
energy and dark matter leads to stable attractor solutions at late
times and the big rip is avoided in the parameter space \cite{Guo}.
Also, it was argued that there are certain classes of unified dark
energy models stable against perturbations, in which cosmic dooms
day  can be avoided \cite{pedro1}. Moreover in scalar tensor
theories, quantum gravity effects may prevent (or, at least, delay
or soften) the cosmic doomsday catastrophe associated with the
phantom \cite{Elizalde,Nojiri2}. Also in Gauss Bonnet gravity theory
and loop quantum cosmology, the big rip occurrence is also avoided
\cite{Nojiri1,Samart}. In order to avoid the big rip with phantom
matter, it is sufficient to have a phantom scalar field with a
potential bounded above by some positive constant \cite{fara}. It is
also suggested that phantom dark energy with $\omega <-1$ can
effectively ameliorate the coincidence problem (i.e.\ why does the
observable universe begin the accelerated expansion so recently and
that why are we living in an epoch in which the dark energy and the
matter energy density are comparable?) \cite{Scherrer,Campo}. In
another model using vector like dark energy with a background of
perfect fluid, it is demonstrated that the cosmic coincidence
problem is fairly solved \cite{Wei}.

The fate of a stationary uncharged black hole in the phantom energy
dominated universe was investigated by Babichev et al \cite{babi}.
The phantom energy was assumed to be a perfect fluid. The phantom
energy was allowed to fall onto the black hole horizon only in the
radial direction. It was concluded that black hole will lose mass
steadily due to phantom energy accretion and disappear near the big
rip. We here adopt their procedure for a static, stationary and
charged black hole. The gravitational units are chosen for this
work.

The paper is organized as follows: In the second section, we have
explained the relativistic model of accretion onto a charged black
hole and obtained the black hole mass loss rate. In third section,
we have determined the critical points of accretion model and have
analyzed the dynamics about these points. Finally we conclude our
paper.

\section{Accretion onto Charged Black Hole}

We consider a static and spherically symmetric black hole of mass
$M$ having electric charge $e$, so-called Reissner-Nordstr\"{o}m
(RN), specified by the line element
\begin{equation}
ds^{2}=f(r)dt^{2}-f(r)^{-1}dr^{2}-r^{2}(d\theta ^{2}+\sin^{2} \theta
d\varphi ^{2}),  \label{1}
\end{equation}
where
\begin{equation}
f(r)=1-\frac{2M}{r}+\frac{e^{2}}{r^{2}}.
\end{equation}
If $e^2>M^2$ then the metric is non-singular everywhere except at
the curvature or the irremovable singularity at $r=0.$ Also if
$e^2\leq M^2$ then the function $f(r)$ has two real roots given by
\begin{equation}
r_{h\pm}=M\pm\sqrt{M^2-e^2}.
\end{equation}
These roots physically represent the apparent horizons of the RN
black hole. The two horizons are termed the inner $r_{h-}$ and the
outer $r_{h+}$. The outer horizon is effectively called the
\textit{event horizon} while the inner one is called the
\textit{cauchy horizon} of the black hole. The metric (3) is then
regular in the regions specified by the inequalities:
$\infty>r>r_{h+}$, $r_{h+}>r>r_{h-}$ and $r_{h-}>r>0$. Note that if
$e^2=M^2$, then it represents an \textit{extreme} RN black hole
while if $e^2>M^2$, it yields a \textit{naked singularity} at $r=0$
\cite{hawk,poi}.

The phantom energy is assumed to be a perfect fluid specified by the
stress energy tensor
\begin{equation}
T_{\mu \nu }=(\rho +p)u_{\mu }u_{\nu }-pg_{\mu \nu }.  \label{2}
\end{equation}
Here $p$ is the pressure and $\rho $ is the energy density of the
phantom energy. Also $u^{\mu }=(u^{t}(r),u^{r}(r),0,0)$ is the four
velocity of the phantom fluid which satisfies the normalization
condition $u^{\mu }u_{\mu }=-1.$ We assume that the in-falling
phantom fluid does not disturb the global spherical symmetry of the
black hole. Further the energy-momentum conservation $T_{;\nu }^{\mu
\nu }=0=T^{tr}_{;r}$ gives
\begin{equation}
ur^{2}M^{-2}(\rho
+p)\sqrt{1-\frac{2M}{r}+\frac{e^{2}}{r^{2}}+u^{2}}=C_{1,}  \label{3}
\end{equation}
where $u^{r}=u=dr/ds$ is the radial component of the velocity four
vector and $C_{1}$ is a constant of integration. For inward flow, we
will take $u<0$. Moreover, the second constant of motion is obtained
by projecting the energy conservation equation onto the velocity
four vector as $u_{\mu }T_{;\nu }^{\mu \nu }=0,$ which yields
\begin{equation}
ur^{2}M^{-2}\exp\left[{ \int\limits_{\rho _{\infty }}^{\rho
_{h}}\frac{d\rho ^{\prime }}{\rho ^{\prime }+p(\rho ^{\prime
})}}\right]=-A.  \label{4}
\end{equation}
Here $A$ is a constant of integration. Above $\rho _{h}$ and $\rho
_{\infty }$ are the energy densities of the phantom energy at the
horizon and at infinity respectively. From Eqs. (7) and (8) we have
\begin{equation}
(\rho +p)\sqrt{1-\frac{2M}{r}+\frac{e^{2}}{r^{2}}+u^{2}}\exp
\left[-\int\limits_{\rho _{\infty }}^{\rho _{h}}\frac{d\rho ^{\prime
}}{\rho ^{\prime }+p(\rho ^{\prime })}\right]=C_{2},  \label{5}
\end{equation}
where $C_{2}=-C_{1}/A=\rho _{\infty }+p(\rho _{\infty })$. In order
to calculate the rate of change of mass of black hole we integrate
the flux of the fluid over the entire cross-section of the event
horizon as
\begin{equation}
\dot{M}=\oint T_{t}^{r}dS,  \label{6}
\end{equation}
where $T_{t}^{r}$ determines the momentum density in the radial direction and $dS=\sqrt{%
-g}d\theta d\varphi$ is the surface element of the horizon, where
$g$ is the determinant of the metric. From Eqs. (7 - 10), we get
\begin{equation}
\frac{dM}{dt}=4\pi AM^{2}(\rho_\infty(t) +p_\infty(t)),  \label{7}
\end{equation}
which clearly demonstrates that mass of black hole decreases if
$\rho_\infty +p_\infty<0.$ Note that Eq. (11) can be solved for any
equation of state of the form $p=p(\rho)$ or in particular
$p=\omega\rho$. In general, Eq. (11) holds for all $\rho$ and $p$
violating the dominant energy condition, thus we can write
\cite{diaz,moru}
\begin{equation}
\frac{dM}{dt}=4\pi AM^{2}(\rho(t) +p(t)).  \label{7}
\end{equation}
In the astrophysical context, the mass of black hole is a dynamic
quantity. The mass increases by the accretion of matter and can
decrease by the accretion of the phantom energy. Since we are not
incorporating matter in our model, the mass of black hole will
decrease correspondingly.

\section{Critical Accretion}

We are interested only in those solutions that pass through the
critical point as these correspond to the material falling into the
black hole with monotonically increasing speed. The falling fluid
can exhibit variety of behaviors near the critical point of
accretion, close to the compact object. For instance, for a given
critical point $r=r_c$, we have the following possibilities
\cite{padmanabhan}: (a) $u^2=c_s^2$ at $r=r_c$, $u^2\rightarrow0$ as
$r\rightarrow\infty$, $u^2<c_s^2$ for $r>r_c$ and $u^2>c_s^2$ for
$r<r_c$. Thus for large distance, the speed of flow becomes
negligible (subsonic), at the critical point it is sonic, while the
flow becomes supersonic for very small $r$. Other solutions for the
flow near $r_c$ are not of much interest due to their
impracticality, like (b) $u^2<c_s^2$ for all values of $r$ and (c)
$u^2>c_s^2$ for all values of $r$. Solutions (b) and (c) are not
realistic since they describe both subsonic and super-sonic flows
for all $r$. Similarly, (d) $u^2=c_s^2$ for all values of $r>r_c$
and (e) $u^2=c_s^2$ for all values of $r<r_c$. Last two solutions
are also useless since they give same value of speed at a given $r$.
Hence from this discussion, we see that solution (a) is the only
physically motivated, near the critical point. To determine the
critical points of accretion we shall adopt the procedure as
specified in Michel \cite{Mich}. The equation of mass flux
$J_{;r}^{r}=0$ gives
\begin{equation}
\rho ur^{2}=k_{1},  \label{13}
\end{equation}
where $k_{1}$ is constant of integration. Dividing and then\
squaring Eqs. (7) and (13) give
\begin{equation}
\left(\frac{\rho +p}{\rho
}\right)^{2}\left(1-\frac{2M}{r}+\frac{e^{2}}{r^{2}}+u^{2}\right)=\left(\frac{C_{1}}{k_{1}}\right)^{2}=C_{3}.
\label{14}
\end{equation}
Here $C_3$ is a positive constant. Differentiation of Eqs. (13) and
(14) and then elimination of $d\rho $ gives
\begin{equation}
\frac{du}{u}\left[2V^2-\frac{\frac{M}{r}-\frac{e^2}{r^2}}{1-\frac{2M}{r}+\frac{e^2}{r^2}+u^2}\right]
+\frac{dr}{r}\left[V^2-\frac{u^2}{1-\frac{2M}{r}+\frac{e^2}{r^2}+u^2}\right]=0,
\end{equation}
or
\begin{equation}
\frac{du}{dr}=-\frac{u}{r}\frac{\left[V^2-\frac{u^2}{1-\frac{2M}{r}+\frac{e^2}{r^2}+u^2}\right]}
{\left[2V^2-\frac{\frac{M}{r}-\frac{e^2}{r^2}}{1-\frac{2M}{r}+\frac{e^2}{r^2}+u^2}\right]}=\frac{N}{D}.
\end{equation}
where
\begin{equation}
V^{2}\equiv\frac{d\ln (\rho +p)}{d\ln \rho }-1.
\end{equation}
We have assumed that the flow is smooth at all points of spacetime,
however if at any point the denominator $D$ vanishes then the
numerator $N$ must also vanish at that point. Mathematically this
point is called the \textit{critical point} of the flow
\cite{sandip}. Equating the denominator $D$ and numerator $N$ to
zero, we can get the so-called \textit{critical point conditions}
given by
\begin{equation}
u_{c}^{2}=\frac{Mr_{c}-e^{2}}{2r_{c}^{2}}, \label{16}
\end{equation}
and
\begin{equation}
V_{c}^{2}=\frac{Mr_{c}-e^{2}}{2r_{c}^{2}-3Mr_{c}+e^{2}}. \label{17}
\end{equation}
Note that by choosing $e=0$ in the above equations, we can retrieve
the results for the accretion of fluid onto a Schwarzschild black
hole \cite{Mich}. All the quantities with subscript $c$ are defined
at the critical point correspondingly. Physically, the critical
points represent the \textit{sonic point} of the flow i.e. the point
where the speed of flow becomes equal to the speed of sound,
$u_c^2=c_s^2$ or the corresponding Mach number $M_c=1$. This
transition may occur from the initial subsonic to the supersonic or
trans-sonic speeds. For any spherically symmetric spacetime, a
surface where every point is a sonic point is called a \textit{sound
horizon} which itself will be spherical. Any perturbation or
disturbance generated in the flow inside the sound horizon ($r<r_c$)
is eventually pulled towards the black hole singularity and hence
cannot escape to infinity.

It can be seen that the speed of sound (squared) $c_s^2=\partial p
/\partial\rho$ has no physical meaning if the EoS parameter
$\omega<0$ (in $p=\omega\rho$). Thus it will apparently make the
exotic cosmic fluids like the cosmological constant, quintessence
and the phantom energy unstable that can not be accreted onto the
black hole. In order to avoid this problem, Babichev et al
\cite{babi1} introduced a non-homogeneous linear equation of state
(nEoS) given by $p=\alpha(\rho-\rho_o)$, where the constants
$\alpha$ and $\rho_o$ are free parameters. The nEoS can describe
both hydrodynamically stable ($\alpha>0$) and unstable ($\alpha<0$)
fluids. The parameter $\omega$ is related to the nEoS as
$\omega=\alpha(\rho-\rho_o)/\rho$. Notice that $\omega<0$
corresponds to $\alpha>0$ and $\rho>\rho_o$, thus making the phantom
energy as hydrodynamically stable fluid. Therefore the speed of
sound $c_s$ is now well-defined with the nEoS for the phantom
energy. Hence, the phantom energy can fall onto the RN black hole
and can reduce the black hole mass. Since phantom energy reduces
only mass and not charge, a stage is reached when the the cosmic
censorship conjecture becomes violated i.e. $e>m$, the so-called
emergence of a naked singularity.

Now physically acceptable solution of Eq. (16) is obtained if
$u_c^2>0$ and $V_c^2>0$, hence we get
\begin{equation}
2r_{c}^{2}-3Mr_{c}+e^{2}\geq0,
\end{equation}
and
\begin{equation}
Mr_{c}-e^{2}\geq0.
\end{equation}
Eq. (20) can be factorized as
\begin{equation}
2r_{c}^{2}-3Mr_{c}+e^{2}=(r_c-r_{c+})(r_c-r_{c-})\geq0,
\end{equation}
where
\begin{equation}
r_{c\pm}=\frac{1}{4}(3M\pm\sqrt{9M^2-8e^2}),
\end{equation}
which are positive satisfying $r_{c+}>r_{c-}>0$. In general, for
$e\leq m$, the inner critical point will lie between $r_{h-}\leq
r_{c-}\leq r_{h+}$, while the outer one will satisfy $r_{c+}\geq
r_{h+}$. It is obvious that these roots will be real valued if
$9M^2-8e^2\geq0$ or
\begin{equation}
\frac{M^2}{e^2}\geq\frac{8}{9}.
\end{equation}
These roots physically represent the locations of the critical or
sonic points of the flow near the black hole. Notice that both mass
and charge have same dimension of length, therefore all the
inequalities here and below represent dimensionless ratios. From Eq.
(22), we can see that these critical points specify two regions for
the flow: (1) $r_c>r_{c+}$ or (2) $0<r_c<r_{c-}$. We shall now solve
Eq. (19) using (21) and then deduce a condition for the black hole
mass and charge.

To get solutions about the critical points, we substitute $r_{c\pm}$
in Eq. (21). For $r_{c+}$, Eq. (21) gives
\begin{equation}
M\sqrt{9M^2-8e^2}\geq4e^2-3M^2,
\end{equation}
which is satisfied if
\begin{equation}
\frac{M^2}{e^2}\leq1,
\end{equation}
and
\begin{equation}
\frac{M^2}{e^2}<\frac{4}{3}.
\end{equation}
A comparison of inequalities (24), (26) and (27) imply
\begin{equation}
\frac{8}{9}\leq\frac{M^2}{e^2}<\frac{4}{3}.
\end{equation}
Thus accretion through $r_{c+}$ is possible if the above inequality
(28) is satisfied. It encompasses the two types of black holes in
itself: regular and the extreme RN black hole. Interestingly, the
naked singularity also falls within the prescribed limits. Thus for
all these spacetimes, the accretion is allowed through the critical
point $r_{c+}$. We stress here that using $e=0$ in the inequality
(28) to retrieve same condition for the Schwarzschild black hole can
be misleading. The inequality is deduced using the outer apparent
horizon and a critical point. Since Schwarzschild black hole
($e\rightarrow0$) possesses unique horizon and the critical point,
the above inequality cannot be reduced for an uncharged black hole.

Now we consider case (2) when $0<r_c<r_{c-}$. Substitution of
$r_{c-}$ in Eq. (21) gives
\begin{equation}
M\sqrt{9M^2-8e^2}\leq3M^2-4e^2.
\end{equation}
If $3M^2-4e^2<0$ then Eq. (29) does not yield any solution. So we
need $3M^2-4e^2>0$ which yields
\begin{equation}
\frac{M^2}{e^2}>\frac{4}{3},
\end{equation}
Further inequality (29) is satisfied if
\begin{equation}
\frac{M^2}{e^2}<1.
\end{equation}
Since Eqs. (30) and (31) are mutually inconsistent, there is no
solution for $r_c$ in case (2). Thus accretion is not possible
through $r_{c-}$.

Since the mass of black hole is decreasing by the accretion of
phantom energy (see Eq. 12), it implies that at least one critical
point must exist for the fluid flow, which is specified by $r_{c+}$.
This critical point yields the mass to charge ratio of the black
hole in the range specified by (28) which allows that accretion onto
all charged spherically symmetric black holes.

\section{Conclusion}

We have analyzed the effects of accretion of phantom energy onto a
charged black hole. The analysis is performed using two critical
points $r_{c\pm}$. It turns out that accretion is possible only
through $r_{c+}$ which yields a constraint on the mass to charge
ratio given by Eq. (28). This expression incorporates both extremal
and non-extremal black holes. Thus all charged black holes will
diminish near the big rip. Apparently this condition predicts the
existence of large charges onto black holes, although
astrophysically no such evidence has been successfully deduced from
the observations. In theory, the existence of large charges onto
black holes is consistently deduced by the general theory of
relativity. It needs to be stressed that there is no analogous
condition for the Schwarzschild black hole ($e=0$). This analysis
can be extended for a rotating charged black hole (so-called
Kerr-Neumann black hole) to get a deeper insight of the accretion
process. This work also serves as the generalization of Michel
\cite{Mich} in terms of the accretion of phantom dark energy onto a
charged black hole.

\subsubsection*{Acknowledgment}
One of us (MJ) would like to thank V. Dokuchaev and E. Babichev for
enlightening discussions during this work. We would also thank
anonymous referees for giving useful comments to improve this work.

\end{document}